\documentclass[twocolumn,secnumarabic,amssymb, nobibnotes, aps, prl]{revtex4-1}
\usepackage{amsmath}
\usepackage[pdftex]{graphicx}
\usepackage{multirow}
\usepackage{array}
\usepackage[FIGTOPCAP]{subfigure}
\usepackage{graphicx}
\usepackage{tabularx}
\usepackage{epsfig}
\usepackage{epstopdf}

\begin{document}

%\title{Bla Bla TASEP Switches}%
\title{Traffic Jams and Shocks of Molecular Motors inside Cellular Protrusions}
\author{I. Pinkoviezky}\author{N. S. \surname{Gov}}%
\email[Corresponding author: ]{nir.gov@weizmann.ac.il}
\affiliation{Department of Chemical Physics, The Weizmann Institute of Science, P.O. Box 26, Rehovot, Israel 76100}

\begin{abstract}
Molecular motors are involved in key transport processes inside
actin-based cellular protrusions. The motors carry cargo proteins to
the protrusion tip which participate in regulating the actin
polymerization, and play a key role in facilitating the growth and
formation of such protrusions. It is observed that the motors
accumulate at the tips of cellular protrusions, and in addition form
aggregates that are found to drift towards the protrusion base at
the rate of actin treadmilling. We present a one-dimensional driven
lattice model, where motors become inactive after delivering their
cargo at the tip, or by loosing their cargo to a cargo-less
neighbor. The results suggest that the experimental observations may
be explained by the formation of traffic jams that form at the tip.
The model is solved using a novel application of mean-field and
shock analysis. We find a new class of shocks that undergo
intermittent collapses, and on average do not obey the
Rankine-Hugoniot relation.
\end{abstract}

\maketitle
%\tableofcontents

The traffic of molecular motors is an example of a non-equilibrium
process
\cite{kolomeisky2007molecular}. In order to describe the traffic of
molecular motors the tools and theories of non-equilibrium
statistical mechanics are useful \cite{chou2011non}. In this letter
we focus on the traffic of molecular motors inside actin-based
cellular protrusions, such as filopodia and stereocilia. These
protrusions are of a few microns in length and fractions of microns
in diameter, and contain a polarized bundle of actin filaments
\cite{mattila2008filopodia}. The actin polymerizes at the protrusion
tip, such that it provides the force for the protrusion initiation,
and treadmills (retrograde flow) at a constant rate when the
protrusion reaches a steady-state shape.

\begin{figure}[b]
\subfigure[]{
\includegraphics[scale=.13]{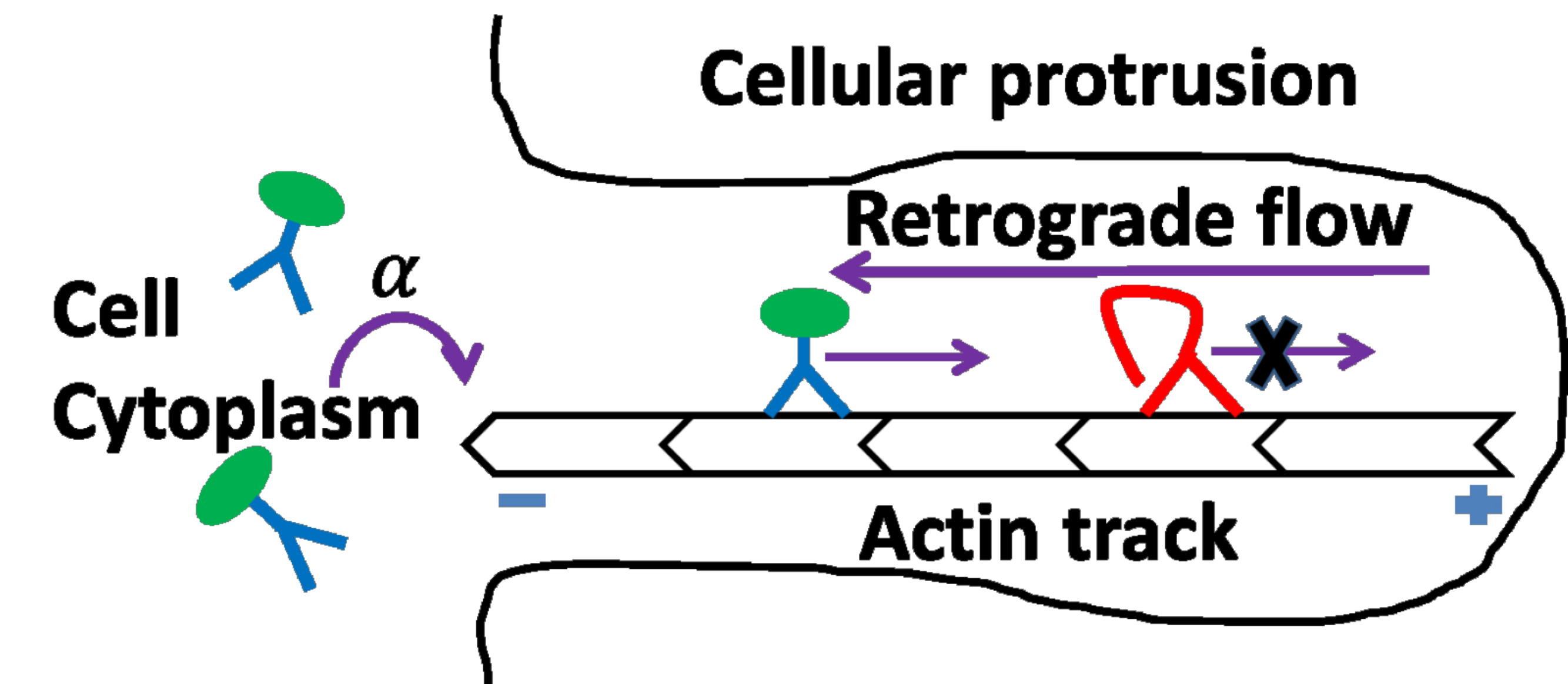}
\label{fig:figure1a}
}
\subfigure[]{

\includegraphics[scale=.12]{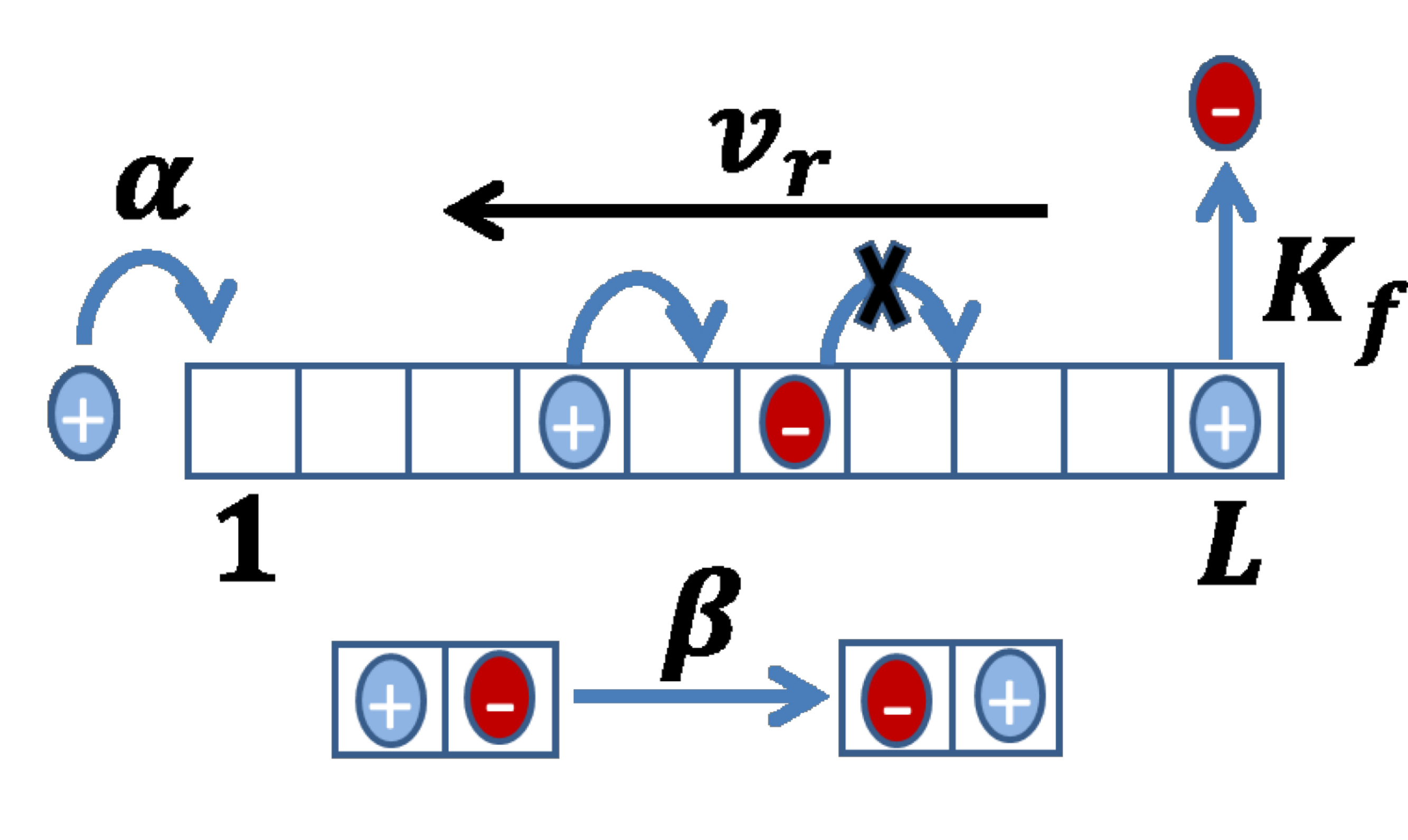}
\label{fig:figure1b}
}
\centering
\subfigure[]{
\includegraphics[scale=.12]{Figure1b-eps-converted-to.pdf}
\label{fig:figure1c} } \caption{(Color online) (a) An illustration
of a cellular protrusion. Myosin motors enter with probability
$\alpha dt$ from the cell's cytoplasm. They move on the actin track.
Green ellipses are cargoes and without it the tail folds to inhibit
the motor domain. (b) The model. Particles enter the
first site with probability $\alpha dt$ an 'switch off' only at the
last site; there is no exit from the left. The track itself moves
backward with probability $v_r dt$. Particles can exchange their
mobility state. (c) Particles interaction. Particles can exchange
cargoes and change their mobility state. } \label{fig:figure1}
\end{figure}

Unconventional myosins bind and move processively toward the
protrusion tip on these filaments, as shown in Fig.
\ref{fig:figure1}a. Experiments analyzing myosin traffic revealed
that motors accumulate over a finite length scale from the tip in
protrusions of different lengths, for example see
\cite{schneider2006new}. A striking phenomenon is seen in a variety
of experiments with different types of myosin motors
\cite{berg2002myosin,belyantseva2005myosin,salles2009myosin}: The
traffic of motors exhibit wave-trains, or "pulses" of motor density,
that move towards the base of the protrusion (opposite to the
motors' motion). The velocity of these aggregates is found to be
close to that of the actin retrograde flow, suggesting that these
are inactive or jammed motors. The theoretical challenge for a
successful model is to explain both the finite length of the
accumulation of motors at the tips of the protrusions, and to
provide a mechanism for the pulse-like counter-propagating
aggregates of motors.

The simplest description of motors along a linear track is in terms
of a Total Asymmetric Exclusion Process (TASEP) \cite{chou2011non}.
We can model a protrusion as a half closed tube, open at its base to
the cell cytoplasm (Fig.\ref{fig:figure1}a). Several works have
dealt with this boundary condition (b.c.) together with
attachment/detachment kinetics of the motors to the tracks
\cite{muller2005molecular,naoz2008protein,zhuravlev2012theory}.
These models find that at steady-state the tubes are practically all
jammed, and longer tubes have longer jammed (accumulation of motors)
regions, in contrast with the observations for cellular protrusions.
In \cite{parmeggiani2003phase} it was shown that a track coupled to
an infinite reservoir produces an accumulation with a fixed length
for different system sizes (see Fig.S1, \cite{SI}). However it is unlikely
that the confined volume of a cellular protrusion can serve as an
infinite reservoir.

Similarly, the properties of the observed pulse-like aggregates of
motors do not fit the traffic jams arising in TASEP. For example, It
seems from experiments that the aggregates originate only at the
protrusion tips, are rather stable while propagating and have low
density regions between them. This is not what happens in TASEP
where jams appear at the high density (HD) phase
\cite{domb2000phase}, and jams appear everywhere. Sparse jams
between free flow regions appear in models of vehicular traffic
\cite{schreckenberg1995discrete}, but they result from the
combination of synchronous update and several particle velocities.
There is no obvious reason why this should apply for molecular
motors traffic.

To explain the observed phenomena we consider a generalization of
TASEP as described in Fig. \ref{fig:figure1}b. Each particle
correspond to a molecular motor, and can be in one of two states:
'inactive' where it is immobile on the actin track, and 'active'
where it can hop (move procssively along the actin filament).
We note that internal degree of freedom was considered in the past
\cite{nishinari2005intracellular,reichenbach2006exclusion,klumpp2008effects,pinkoviezky2013modelling}.
Motivated by experiments and theoretical models, we propose the
following properties for the dynamics of the activity state of the
motors: It was found that several types of molecular motors become
processive only when they are bound to a cargo molecule
\cite{umeki2011phospholipid,manor2012competition}. Since in many
cases the cargo is involved in regulating the actin polymerization
at the protrusion tip
\cite{tokuo2004myosin,salles2009myosin,manor2011regulation}, we
assume that the motors can only detach from the cargo at the tip and
become inactive. When inactive, motors may detach from the actin
filament, or stay attached \cite{umeki2011phospholipid} and drift
towards the protrusion base due to the retrograde flow. Furthermore,
neighboring motors can "steal" the cargo from each other
\cite{manor2012competition}, and this introduces a conservation of
the activity whereby an inactive motor can only become active at the
expense of a motor jammed behind it (Fig.\ref{fig:figure1}c). Such
an interaction is known to produce a robust traffic-jam (condensate)
\cite{mallick1996shocks}. Note that there are other forms of
interactions among motors that result in their inactivation \cite{quintero2010intermolecular}, and
therefore our model may be treated as a coarse-grained description
of more complex set of underlying interactions.

\begin{figure}[t]
\subfigure[]{
\includegraphics[scale=.22]{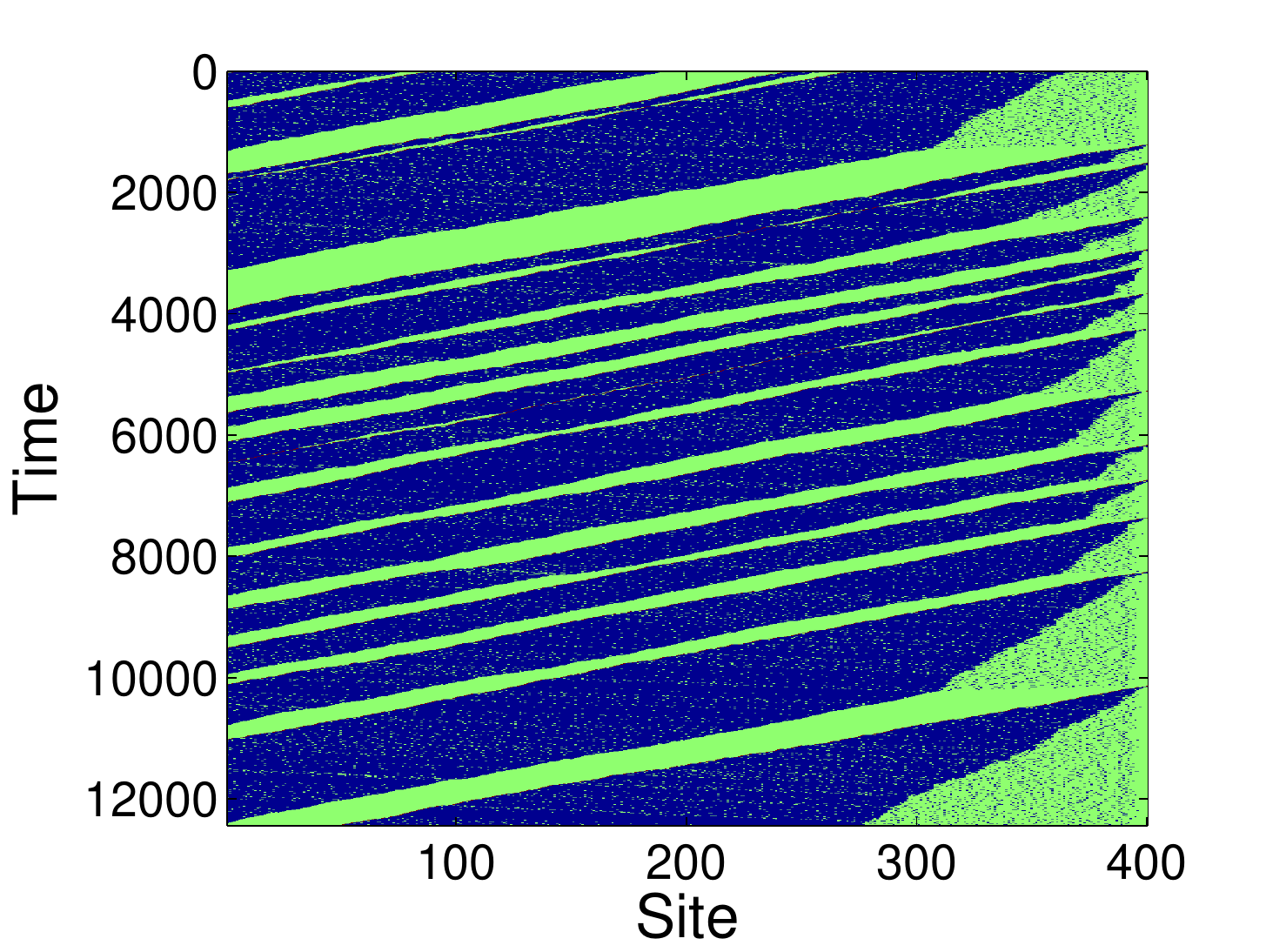}
\label{fig:figure2a}
}
\subfigure[]{

\includegraphics[scale=.065]{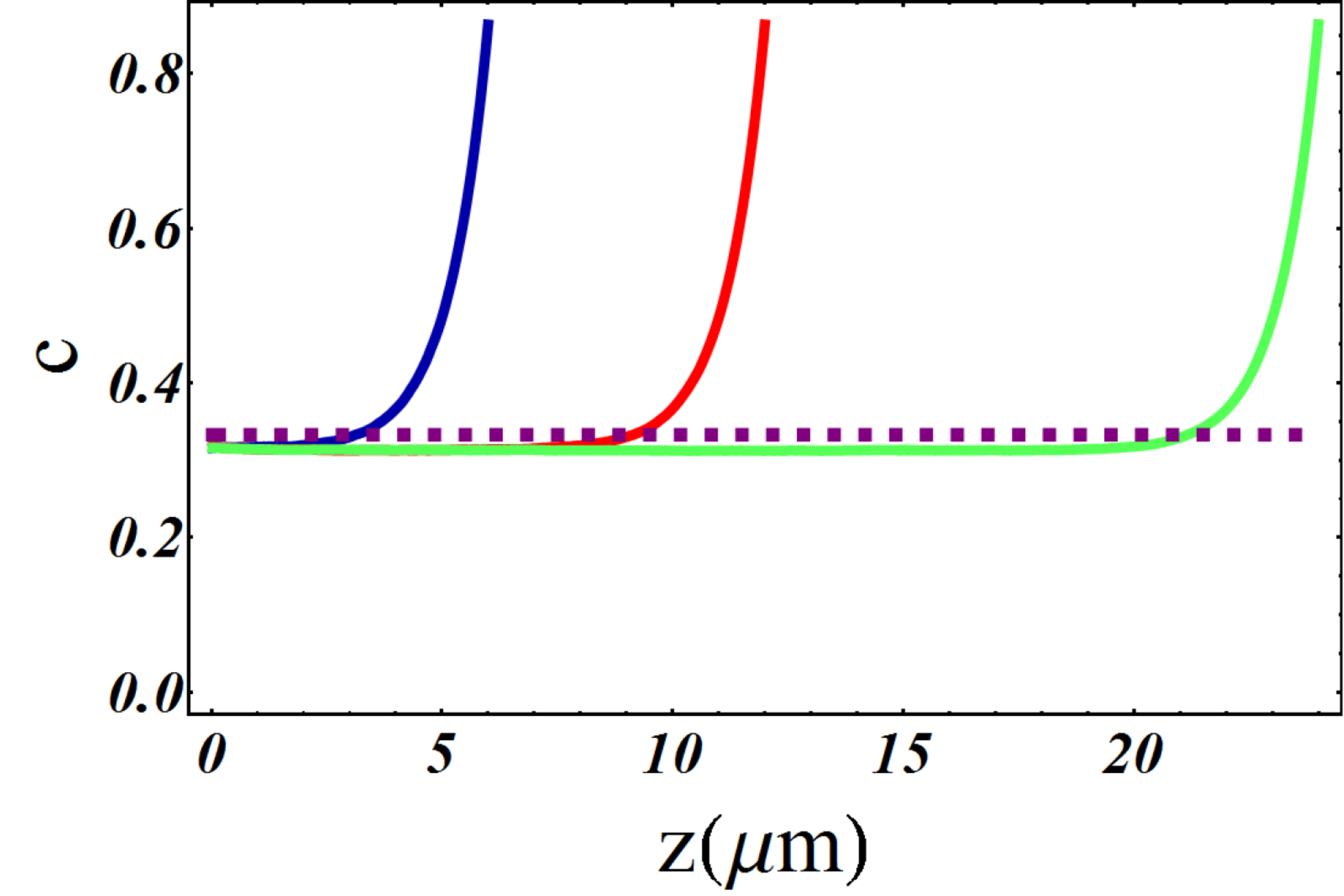}
\label{fig:figure2b}
}

\caption{(Color online) (a) Kymograph. Green (bright) points are
active particles while dark blue is the empty space. (b) Density
profiles for systems of lengths $z=6,12,24\mu m$. Data is for $\alpha=\beta=0.05,v_r=0.1$ and $K_f=0.6/300$. Dashed purple curve is the
theoretical bulk density $c_b$ (Eq.\ref{cx},S4 \cite{SI}).}
\label{fig:figure2}
\end{figure}

The parameters in our model are: $\bf 1)$ $K_{f}dt$ is the
probability for a particle to switch from active to an inactive
state at the tip (at the last site). $\bf 2)$ $\alpha dt$ is the
probability that a particle enters the system at the left boundary
(from the cell cytoplasm). $\bf 3)$ $\beta dt$ is the probability
that an inactive motor followed by an active motor switch their
mobility state (fig. \ref{fig:figure1}c). $\bf 4)$ $v_{r}$ is the
rate at which a new site (actin monomer) is added at the right end
(tip) and simultaneously a site is removed at the left end. The
actin retrograde velocity is therefore $v_r$, and we maintain a
constant overall length of the protrusion which is assumed to be at
steady-state. $\bf 5)$ We normalize the probabilities such that $dt$
is the hopping probability of an active particle.

The process defined above is the open b.c. version of a process with
a single impurity particle on a ring \cite{mallick1996shocks}. In our
process active motors enter from the left while impurity particles
(inactive motors) enter from the right.

We show an example kymograph of the dynamics arising in our model in
Fig. \ref{fig:figure2}a. We find that near the tip there is a region
of accumulation of motors, and large traffic jams are initiated
there. Each traffic jam corresponds to an inactivation event of a
motor at the tip, and these jams keep their size throughout the
system as they move towards the left boundary. When a jam is formed, it transiently depletes the tip
region, which gets refilled shortly afterwards. These properties of
robust aggregates that form near the tip and deplete it, correspond
qualitatively with the experimental observations of myosin-X
\cite{berg2002myosin} and myosin-XV \cite{belyantseva2005myosin} in
filopodia .

\begin{figure}[t]
\subfigure[]{
\includegraphics[scale=.065]{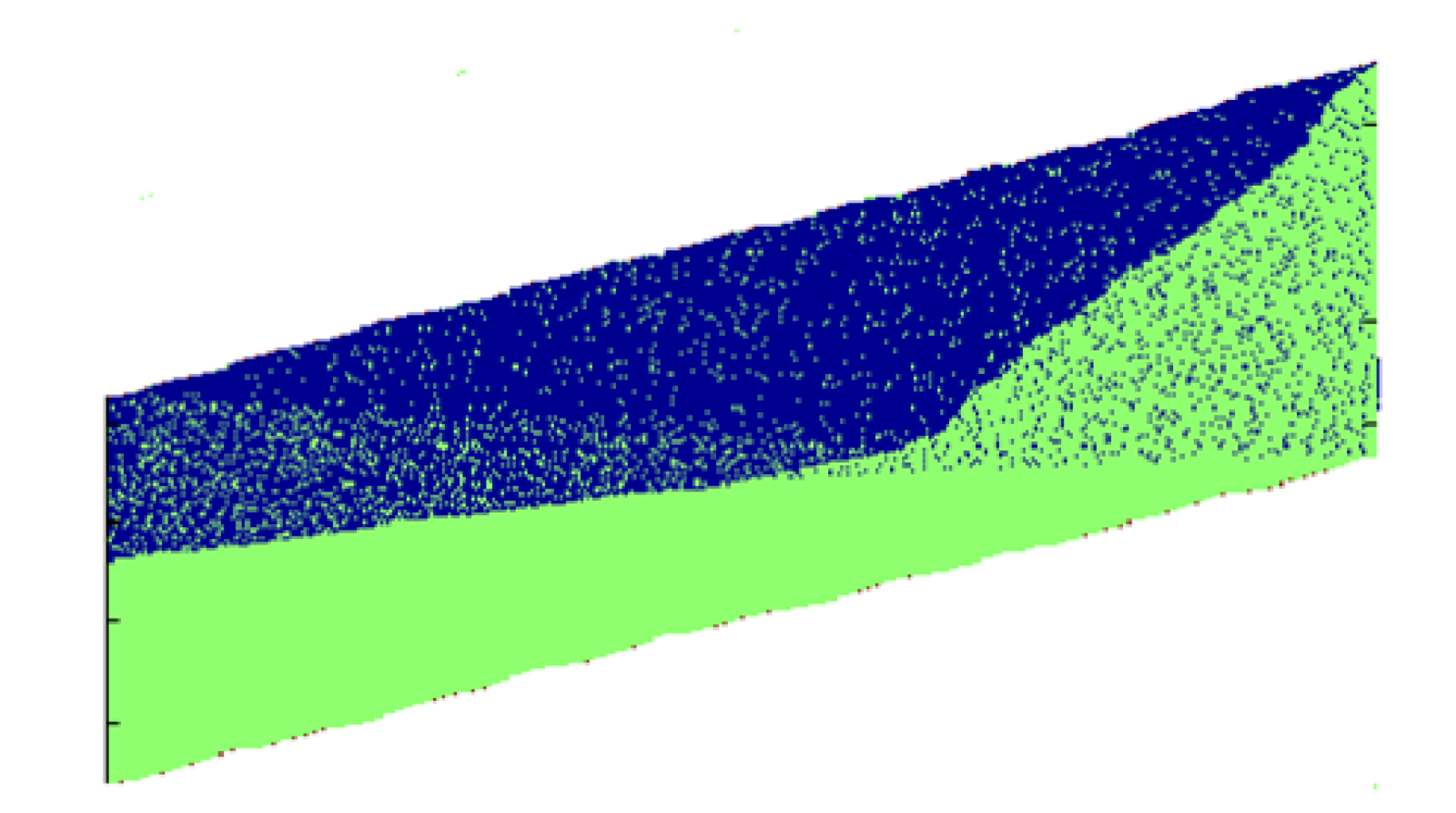}
\label{fig:figure3a}
}
\subfigure[]{

\includegraphics[scale=.15]{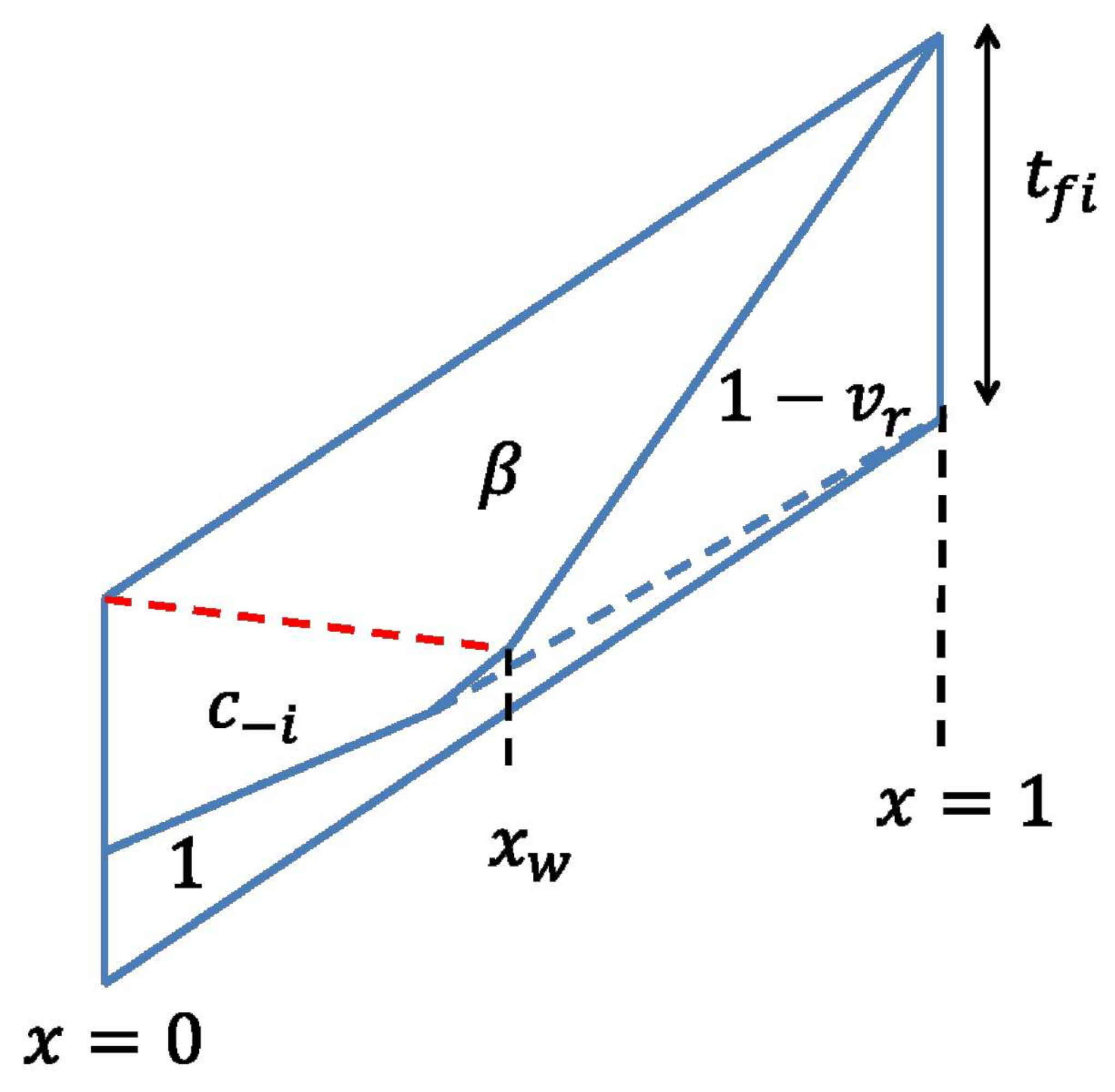}
\label{fig:figure3b}
}

\caption{(Color online) (a) An example of a parallelogram (or slice) for $\alpha=0.2,\beta=0.05,v_r=0.1$ and $\Lambda_f=0.2$. (b) A description of a slice with the different density regions and the slopes connecting them (see \cite{SI} for more examples). In both figures we see $x_w$ (Eq.\ref{xw}) where the densities [$c_-,\beta,1-v_r$] meet.}
\label{fig:figure3}
\end{figure}

We find in Fig. \ref{fig:figure2}b that the accumulation length of
the motors near the tip is independent of the system size. The
spatial extent of each site corresponds to a motor step (i.e.
$30nm$), therefore: $z=i\cdot 0.03\mu m$ where $i$ is the site
number. We stress that without the switching mechanism the track
would be uniformly occupied with density $1-v_{r}$ except for a
shock on the left end. The jams initiated by the inactive particles
determine a finite length of the accumulation region and do not let
it grow to the system size. Such localized accumulations of motors
are observed near the tips of stereocilia of different lengths
\cite{schneider2006new}, and our model suggests that this arises
from the turnover of motors through the formation of jams.

We now turn to analyze the model in detail, at steady state. The
interesting behavior arises in the regime where the inactivation
rate $K_f$ is small, and significant accumulation of motors occurs
at the tip. It is convenient to use the so-called 'mesoscopic
scaling' $K_f=\frac{\Lambda_{f}}{L}$ \cite{reichenbach2006exclusion}
and to use a spatial coordinate $x=\frac{i}{L},0 \leq x \leq 1$,
where $L$ is the number of sites in the system.

We start by comparing the numerical simulations to a naive mean
field approximation (MFA) \cite{derrida1992exact}, which fails, as
demonstrated in Fig. S2 \cite{SI}. The reason for this deviation
lies in the fact that the system separates into regions of different
mean densities and currents. We therefore proceed with a detailed
calculation of the {\em average} concentration profiles of the
motors in our model, which is based on using MFA within each
distinct region. We divide the whole space-time evolution of the
system into parallelogram slices as shown in Fig.
\ref{fig:figure3}a. The $i$'th parallelogram is defined by $t_{fi}$
- the time between two successive inactivation events at the tip.

\begin{figure}[t]

\includegraphics[scale=.25]{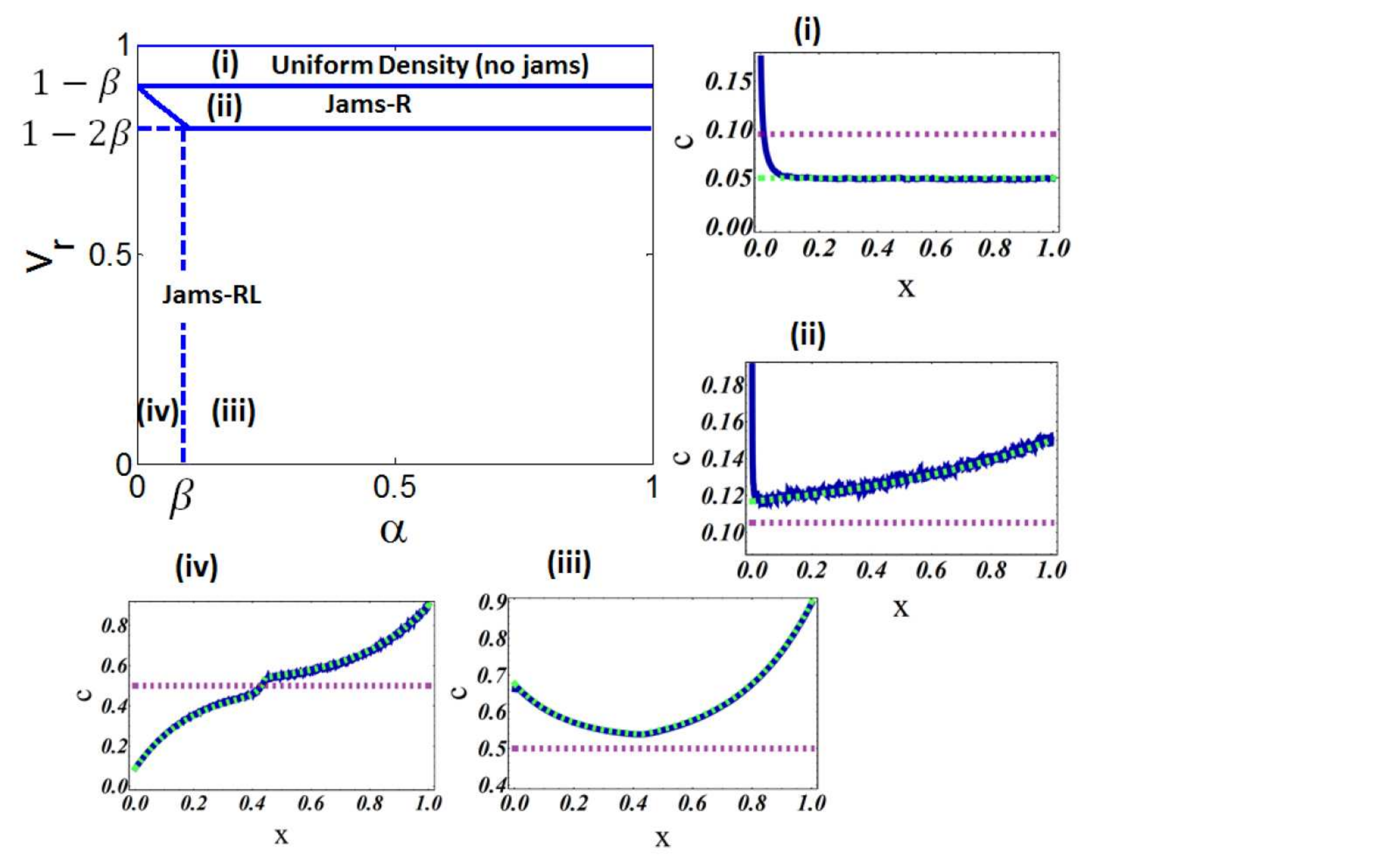}

\caption{(Color online) Phase diagram in $\left(\alpha, v_r \right)
$ space and typical density profiles, using $\beta=0.1$.
$v_r>1-\beta$: Uniform density phase (no jams), with density profile
(i). $1-\beta>v_r>1-\beta-c_-$: Jams-R phase where the behavior is
dominated by the right boundary alone (density profile (ii)).
$1-\beta-c_->v_r$: Jams-LR phase where the behavior is influenced by
both boundaries. The dashed vertical line separates this phase into
two region: (iii) $\alpha>\beta$ and (iv) $\alpha<\beta$. Density
profiles: Blue curves - simulation results; Green dashed curves -
theoretical result of Eq.\ref{cx}; Horizontal dashed purple curves -
bulk density $c_b$.} \label{fig:figure4}
\end{figure}

We find four regions of different average concentrations
(Fig.\ref{fig:figure3}b, for more details see \cite{SI},
Figs.S3,S4): (i) HD region near the tip, with density $1-v_r$, (ii)
jammed regions of density $1$, (iii) free flow regions with density
$\beta$, and (iv) the entrance region near the base with density
$c_-=\min(\alpha,\frac{1-v_r}{2})$. The lines separating these
different regions have slopes that can be calculated according to
the shock velocities \cite{kolomeisky1998phase,SI}. Using these
slopes, we can find the meeting points between the
different regions within the space-time slice. This allows us to
calculate the time-duration that each spatial point spends in a
region of certain density.

One such important triple meeting point is between the regions of
densities $[1-v_r,\beta,c_-]$, which if it exists defines the
meeting between the regions influenced by both boundaries
(Fig.\ref{fig:figure3}). If it does not exist as a real meeting
point, it nevertheless defines the location of the matching between
the left and right solutions. The location of this point is given by
\begin{equation}
x_w=\frac{v_r(1-v_r-\beta-c_-)}{v_r(1-v_r-\beta-c_-)+\beta(1-c_-)}
\label{xw}
\end{equation}

Summing the contributions of each region, we get that the {\em
average} density profile is given by
\small\begin{align} \nonumber
x>x_w :\\ \nonumber
 c(x) &=\frac{\beta}{\beta+v_{r}}+\left(1- v_{r} -\frac{\beta}{\beta+v_{r}}\right)\exp{\left(\frac{x-1}{\xi_{r}}\right)}\\ \nonumber
 x< x_w : \\ \nonumber
 c(x) &= \frac{\beta}{\beta+v_{r}} +\frac{v_r\left( c_{-} - \beta\right)}{(c_{-}+v_r)(\beta+v_r)}\exp{\left( -\frac{x}{\xi_l} \right) }  \\ \nonumber
&+ \left( 1-c_--\frac{c_-}{c_-+v_r}\right)  \cdot \exp{\left(\frac{ \Lambda_{f}(1-v_{r})(c_{-}-\beta)}{c_{-}\beta}(x_w-1)\right)} \cdot\\
& \exp{\left(\frac{\Lambda_{f}(1-v_{r})(1-c_{-})}{c_{-}}(x-1) \right) }
\label{cx}
\end{align}
\normalsize
where
\begin{equation} \xi_{r}
=\frac{\beta}{\Lambda_{f}(1-\beta)(1-v_r)},\xi_l=\frac{v_r
(1-v_r-\beta - c_{-})}{\Lambda_f(1-c_{-})(1-vr)(1-\beta)}
\label{lengths}
\end{equation}
$\xi_l,\xi_r$ are the 'healing' lengths of the left and right
exponentials respectively as shown in Fig.
\ref{fig:figure4}(i)-(iii). The agreement between the simulations and the
calculated density profile (Eq.\ref{cx}) is very good, and improves
for large systems $L\rightarrow \infty$. For comparison, we also
denote the bulk density predicted from the periodic model
\cite{mallick1996shocks} with retrograde flow:
$c_b=\beta/(\beta+v_r)$ \cite{SI}, as seen in Eq.\ref{cx}.

The jam size in the bulk is an exponential random variable with mean
value
\begin{equation}
\left\langle \Omega
\right\rangle=L\frac{\beta}{\Lambda_{f}}\frac{1-v_{r}-\beta}{(1-v_{r})(1-\beta)}
\label{avejam}
\end{equation}
We compare this result with simulation in Fig. S5 in \cite{SI}
and we see that the two agree very well.

The phase diagram of the system, is shown in Fig. \ref{fig:figure4}.
We first note that for $\beta=1-v_r$, both the mean jam size
(Eq.\ref{avejam}) and the exponential accumulation at the tip
(Eq.\ref{cx}) vanishes. For $\beta>1-v_r$ we indeed find that the
system does not exhibit any jams (except for the usual background of
TASEP fluctuations), and the concentration is flat throughout the
system at a value of $1-v_r$ (except for a shock at the left
boundary, Fig.\ref{fig:figure4}i). The system behave as TASEP with
retrograde flow $v_r$ and zero current. This second-order transition
can also be understood in terms of the velocities of the holes
entering the system at the tip \cite{SI}. Note that in the limit
$v_r=0$, inactive particles will get stuck in the first site,
preventing new particles from entering the system. This can be fixed
by changing the b.c., so that the inactive particle is also removed
at the left boundary.

\begin{figure}[t]
\begin{tabular}{cc}
\subfigure[]{ \hspace{-1.cm} \includegraphics[scale=.25]{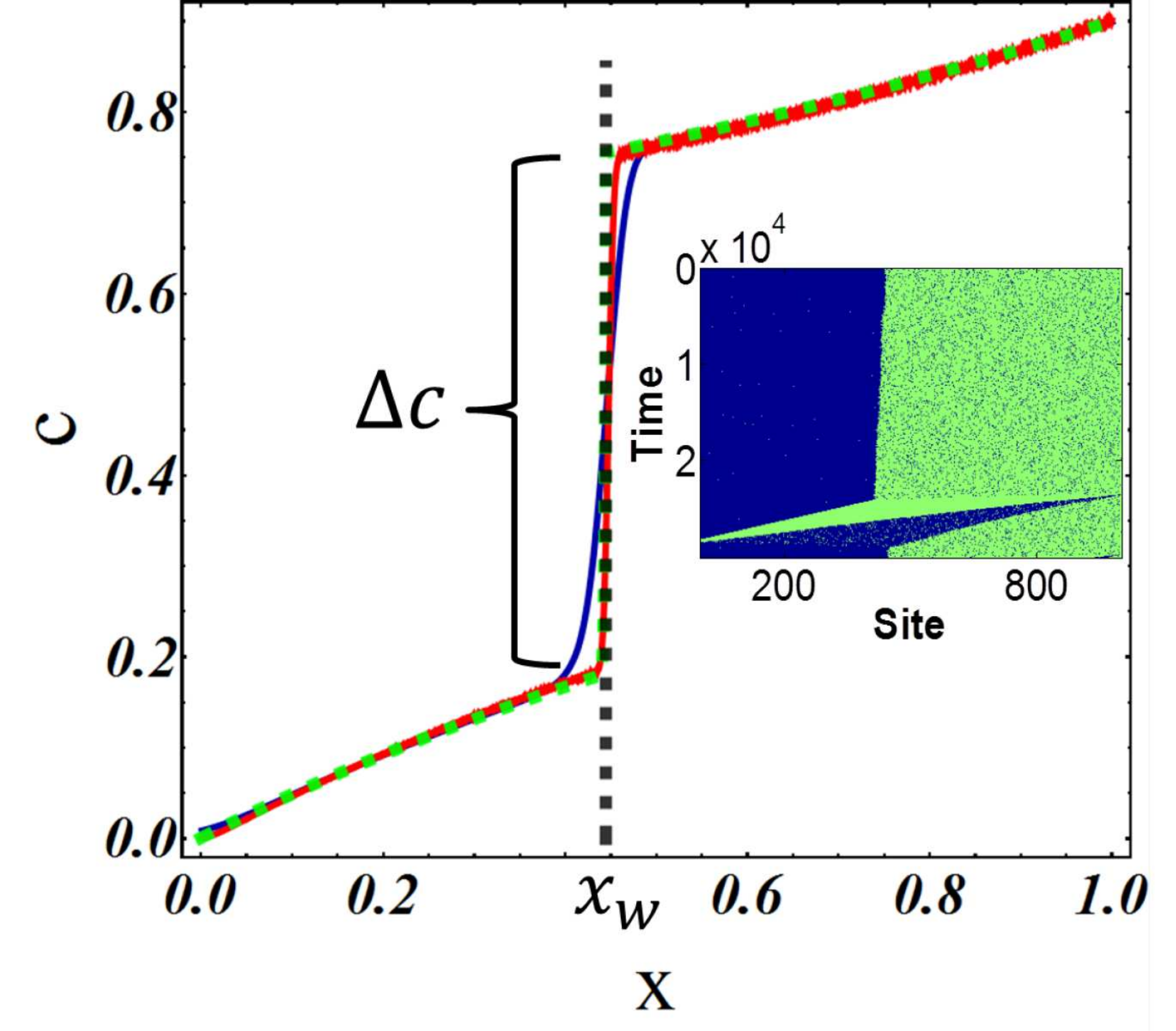}}
\multirow{-3}[-4.1]{*}{\subfigure[]{\includegraphics[scale=.06]{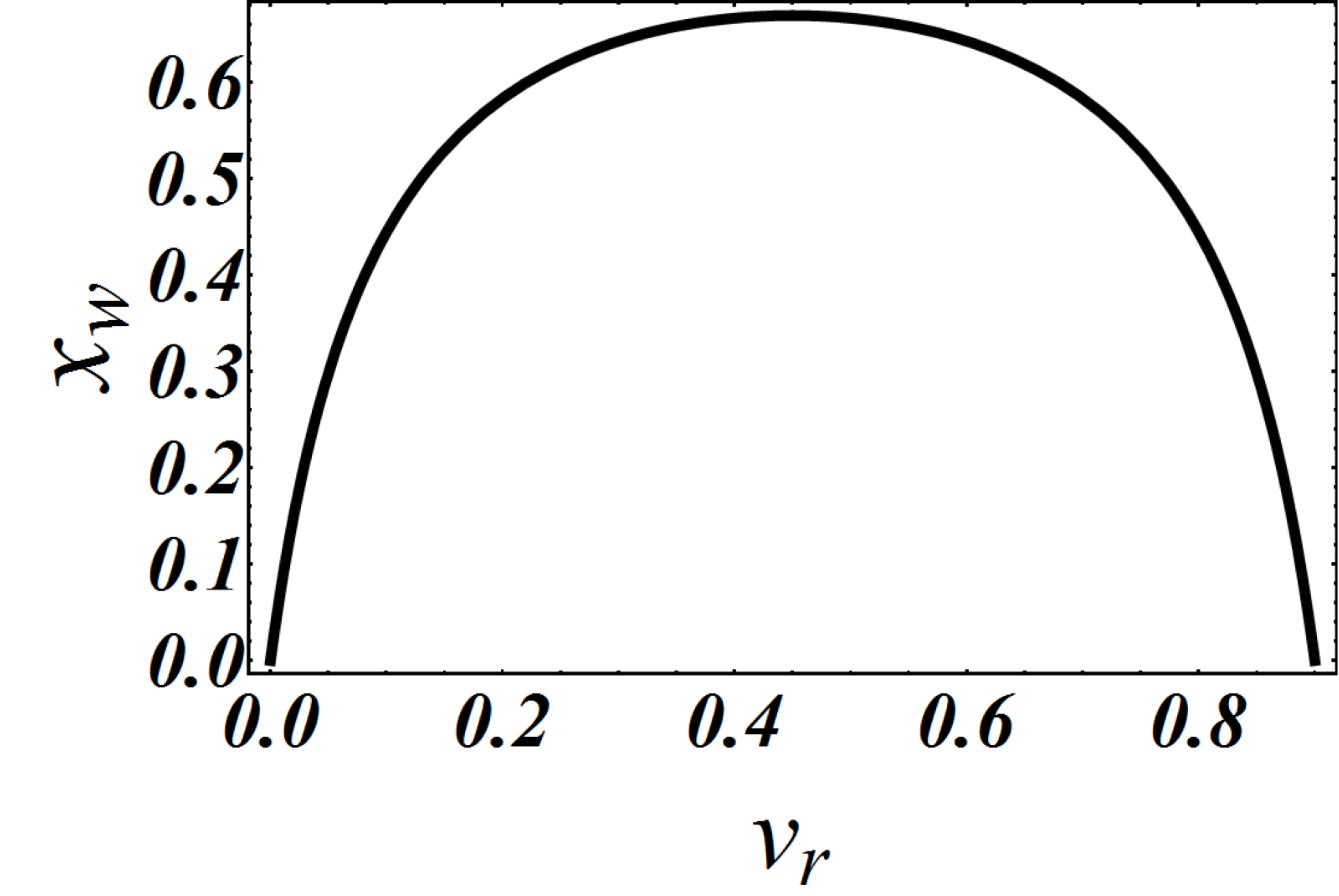}}}
\hspace{-3.5cm}\multirow{-3}[-28.1]{*}{\subfigure[]{\includegraphics[scale=.06]{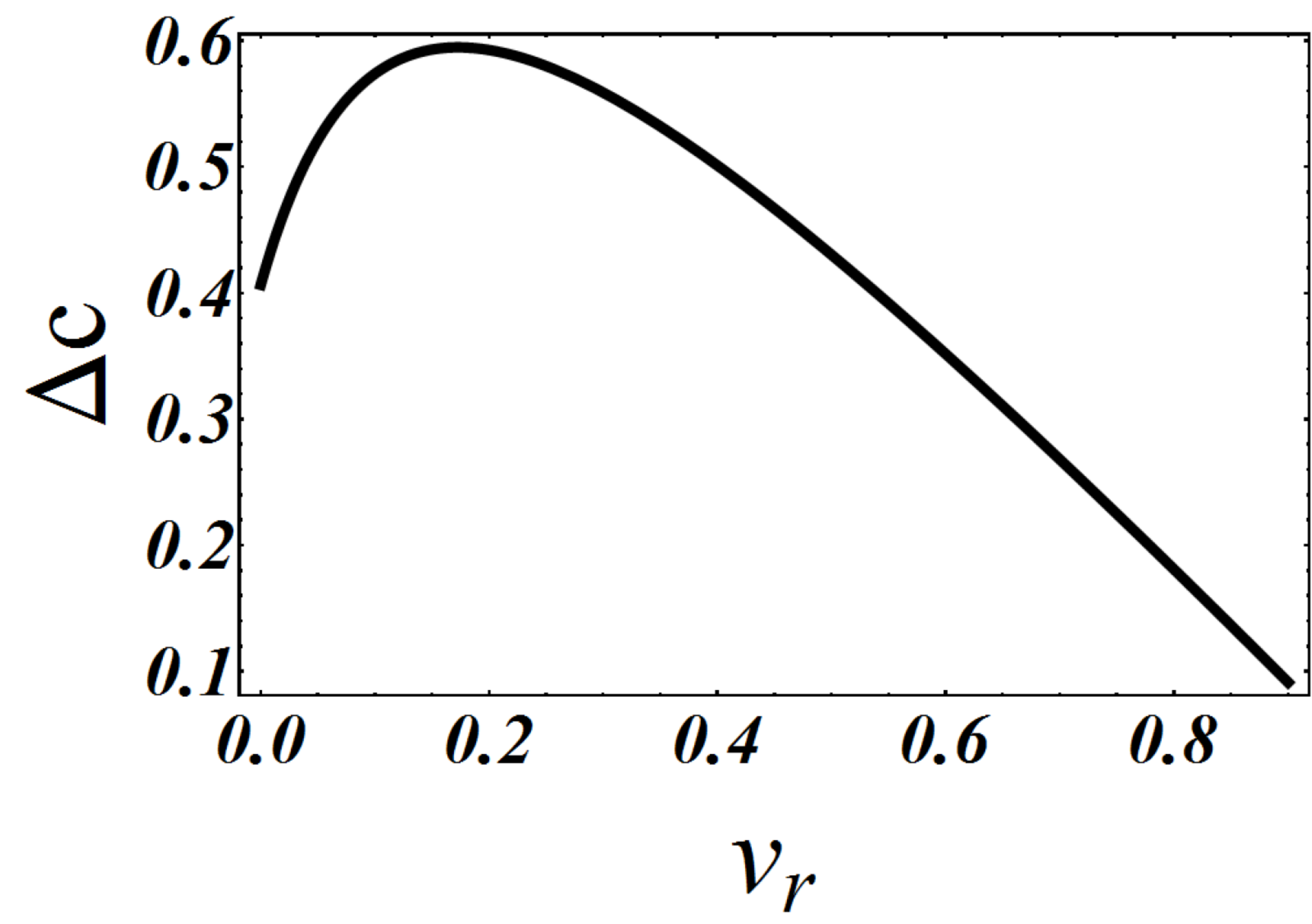}}}
\end{tabular}
\caption{(Color online) System exhibiting a shock: Density profile
for $\alpha=1/L,\beta=0.1,v_r=0.1,\Lambda_f=0.1$ with $L=10^3$ (blue
curve) and $L=10^4$ (red curve). Green dashed line is the
theoretical result of Eq. \ref{cx}. Inset: Kymogrpah of the system.
(b) The shock location $x_w$ from Eq.\ref{xw} and, (c) the density
jump at the shock ($\Delta c$, Eq.S44).} \label{fig:figure5}
\end{figure}

Next, we note that $x_w$ becomes negative when: $1-v_r-c_-<\beta$,
which corresponds to a vanishing of the left exponential in $c(x)$,
and therefore this phase is denoted as Jams-R
(Fig.\ref{fig:figure4}, and profile (ii)). Finally for positive
$x_w$ (denoted as Jams-LR phase in Fig.\ref{fig:figure4}) we find
that: for $\alpha>\beta$ jams grow as they approach the base (near
the left end, Fig.\ref{fig:figure4}iii), while for $\alpha<\beta$
they shrink (Fig.\ref{fig:figure4}iv). This means that for
$\alpha<\beta$ there is a step-like profile, with the step located
at $x_w$. The steepness of the step is given by the exponential
localized at $x_w$ in eq. (\ref{cx}). Taking the limit of
$\alpha\rightarrow0$ by using the scaling $\alpha \sim 1/L$ and
$L\rightarrow \infty$, results in the formation of a shock at $x_w$
(Fig. \ref{fig:figure5}a). We find that the location of the shock in
the system has a reentrant behavior as a function of $v_r$ (Fig.
\ref{fig:figure5}b), while the amplitude of the density
discontinuity ($\Delta c$, Eq.S44) is also non-monotonic (Fig.
\ref{fig:figure5}c).

Unlike previous shocks found in TASEP-like models
\cite{parmeggiani2003phase,reichenbach2006exclusion,hinsch2006bulk,arita2013asymmetric}
the shock we find is only defined for the {\em average}
concentration, while it maintains a dynamic nature: it undergoes
intermittent collapses (kymograph in Fig. \ref{fig:figure5}).
Furthermore, these shocks do not obey the usual Rankine-Hugoniot
relation that follows from naive MFA: $c(x^-_w)=1-c(x^+_w)-v_r$. During the time that no motor is switched
off at the tip, there is an accumulation of particles and a domain
wall fulfilling the shock relation is established:
$c(x^-_w)=0,c(x^+_w)=1-v_r$, so that: $\Delta c|_{x_w} = 1-v_r$. The
probability that such a domain wall exists at any given time is given in Eq.S47, and is simply the fraction of time duration that no jam is initiated
at the tip. %$\exp{\left(-\frac{\Lambda_{f}(1-v_{r})(1-c_-)(1-\beta)}{v_r(1-v_r-\beta-c_-)+\beta(1-c_-)}\right)}$
%\begin{align}
%P_w&=\frac{c(x_{w}^{+})-c(x_{w}^{-})}{1-v_r}=\\ \nonumber
%&\exp{\left(-\frac{\Lambda_{f}(1-v_{r})(1-c_-)(1-\beta)}{v_r(1-v_r-\beta-c_-)+\beta(1-c_-)}\right)}
%\end{align} 

{\em Discussion.} Our model is able to reproduce two experimental
observations of molecular motors in actin-based cellular
protrusions, namely the finite accumulation length of the motors at
the protrusions' tips, and the formation of backward-moving
aggregates of motors from the tip to the base. In our model these
phenomena are linked and both arise from the random process of
traffic-jam initiation at the tip, followed by the "relay-race"-like
transport of the cargo between the motors. Note however, that this
mechanism is maintaining individual motors near the tip, since only
one motor is recycled back to the cytoplasm per traffic jam. Since
there are multiple parallel actin tracks inside a real protrusion,
we expect the turnover of motors to be more efficient than our
one-dimensional model suggests. The effects of such parallel tracks,
as well as those of attachment/detachment kinetics will be explored
in a future elaboration of this work.

Since the cargo carried by the motors is often involved in enhancing
the actin polymerization at the tip \cite{tokuo2004myosin,salles2009myosin,manor2011regulation}, a full treatment of this
system should include a feedback between the system size and
motor/cargo distribution during the growth phase and the
fluctuations around the steady-state length. While in the current
work we considered a fixed geometry, we propose to investigate the
feedback between motors and protrusion length in the future, similar
to \cite{melbinger2012microtubule, sugden2007model,sugden2007dynamically}.

From the theory point of view we introduce here a model that has
several unique features, compared to previous models of molecular
motors on cytoskeletal tracks \cite{lipowsky2001random,johann2012length,
melbinger2012microtubule, sugden2007model}: (i) We find that the MFA
fails, while it works well in separated domains of the system that
are connected through shocks, (ii) we find that while there is a
steady-state {\em average} density, the spatio-temporal behavior of
the system is inherently dynamic exhibiting large fluctuations,
(iii) a discontinuity (shock) can appear in the {\em average}
steady-state density profile, but this shock is inherently dynamic
and undergoes intermittent collapses.

Finally, we can make several qualitative predictions. Both the actin
polymerization rate ($v_r$) and the influx of motors ($\alpha$) may
be modified in experiments, and therefore the phase diagram of fig.
\ref{fig:figure4} explored. We predict that increasing the
retrograde flow will result in a decrease of the average jam size,
as follows from eq. (\ref{avejam}). The parameters
$K_f,\beta$ are controlled by the cargo affinity to the motors: By
modifying $\beta$ the system will change its phase according to Fig.
\ref{fig:figure4}a, and through $K_f$ the average size of the jams can
be manipulated (eq. \ref{avejam}).

\begin{acknowledgments}

We gratefully acknowledge funding from the ISF (grant no. 580/12).
This research is made possible in part by the historic generosity of
the Harold Perlman Family. We thank Kirone Mallick and Tridib Sadhu
for useful discussions.

\end{acknowledgments}
\bibliographystyle{apsrev4-1}
\bibliography{mylib}
\end{document}